\documentclass[aps,11pt,final,notitlepage,oneside,onecolumn,nobibnotes,nofootinbib,superscriptaddress,showpacs,centertags,showkeys,amsmath,amssymb]{revtex4}
 \usepackage[dvips,final]{graphicx}
  \usepackage{dcolumn} 
   \usepackage{pifont} 

\graphicspath{{./Figs-BMPS/}}

\begin{document}

\title{Dispersion relations and QCD factorization in hard reactions}

\author{\underline{I.V.~Anikin}}
\affiliation{Bogoliubov Lab. of Theoretical Physics, JINR, 141980 Dubna, Russia}

\author{O.V.~Teryaev}
\affiliation{Bogoliubov Lab. of Theoretical Physics, JINR, 141980 Dubna, Russia}

\begin{abstract}
We study analytical properties of the hard exclusive
process amplitudes. 
We found that QCD factorization 
for deeply virtual Compton scattering and hard exclusive vector meson 
production results in the subtracted dispersion 
relation with the subtraction constant determined by the Polyakov-Weiss $D$-term. 
\end{abstract}


\maketitle

\section{Introduction}

Hard exclusive reactions described by the Generalized Parton 
Distributions (GPDs) 
are the subject of extensive theoretical and experimental studies \cite{DM}-\cite{belrad}. 
The analytical properties of deeply virtual Compton scattering (DVCS) and hard exclusive vector meson 
production (VMP) amplitudes constitute the important aspect of these studies \cite{T-An}-\cite{PolT}. 
The crucial point in application of the relevant 
dispersion relations is a possible ambiguity due to the subtraction constants. 
In this talk we address the problem of dispersion relations and subtractions in the framework 
of the leading order QCD factorization \cite{AT}.        

\paragraph{Some reminders (in a nutshell)}
For an complex function $f(z)$ which is analytic in region $D$, the Cauchy formula reads
\begin{eqnarray}
f(z)=\frac{1}{2\pi i} \int\limits_{{\bf C}}d\omega \frac{f(\omega)}{\omega- z},
\nonumber
\end{eqnarray}
where contour $C$ restricts some domain $D$. Choosing the contour which includes the infinite semi-circle and
the line $\omega=x^\prime +i\epsilon$  and assuming a ``good" behaviour at infinity, {\it i.e.} 
$\left| f(\omega)\right| < |\omega|^{-1}$ at $\left| \omega \right| \to \infty$,
one gets $(\epsilon > \epsilon^\prime)$
\begin{eqnarray}
\lim_{\epsilon\to 0}f(x+i\epsilon)=\lim_{\{\epsilon,\epsilon^\prime\}\to 0}
\frac{1}{2\pi i} \int\limits^{\infty}_{-\infty}dx^\prime 
\frac{f(x^\prime+i\epsilon^\prime)}{x^\prime - x-i(\epsilon-\epsilon^\prime)}
\nonumber
\end{eqnarray} 
Let us now put the case that
$\left| f(\omega)\right| \sim const$ at $\left| \omega \right| \to \infty$. In this case,
the Cauchy formula gives us
\begin{eqnarray}
f(x+i\epsilon)=
\frac{1}{2\pi i} \int\limits^{\infty}_{-\infty}dx^\prime 
\frac{f(x^\prime+i\epsilon^\prime)}{x^\prime - x-i(\epsilon-\epsilon^\prime)}
+\Gamma_\infty,
\nonumber
\end{eqnarray}
where $\Gamma_\infty$ denotes the contribution from the integration over the infinite semi-circle.
If $\left| f(\infty)\right| \sim const$,  
$\Gamma_\infty$ is finite. However, the integral over the real axis is infinite.

Let us now make Jordan's lemma to be applicable. We implement the standard trick:
$f(\omega)\rightarrow {\cal F}(\omega)=f(\omega)/\omega$.
Hence, the Cauchy formula reads
\begin{eqnarray}
{\rm Re} {\cal F}(x)=
\frac{v.p.}{\pi} \int\limits^{\infty}_{-\infty}dx^\prime 
\frac{{\rm Im} {\cal F}(x^\prime+i\epsilon^\prime)}{x^\prime - x}.
\nonumber
\end{eqnarray} 
Expressing in the terms of function $f(x)$, one obtains one-subtracted dispersion relations:
\begin{eqnarray}
{\rm Re} f(x)=
\frac{x}{\pi}\,v.p.\, \int\limits^{\infty}_{-\infty}dx^\prime 
\frac{{\rm Im} f(x^\prime+i\epsilon^\prime)}{x^\prime(x^\prime - x)}+{\rm Re} f(0).
\end{eqnarray}

We now focus on the case when the function has a branch point on the real axis.
Then, from the Cauchy theorem, one has
\begin{eqnarray}
{\rm Re} f(\nu)&=&
\frac{v.p.}{\pi}\int\limits^{\infty}_{\nu_{th}}d\nu^\prime 
{\rm Im} f(\nu^\prime+i\epsilon^\prime)
\Biggl[ \frac{1}{\nu^\prime - \nu}+\frac{1}{\nu^\prime + \nu}\Biggr]
+\Gamma_\infty
\nonumber
\end{eqnarray}
with
${\rm Im} f(\nu\pm i\epsilon)={\rm Im} f(-\nu\mp i\epsilon)= - {\rm Im} f(-\nu\pm i\epsilon)$.
Assume that $\left| f(\infty)\right| \sim |\omega|^a$ with $1 < a < 2$ and we are interested in the dispersion 
relations for the symmetric part of $f(\omega)$: $f^{(+)}(\omega)\equiv f(\omega)+f(-\omega)$.
In this case, owing to the symmetry properties, one gets again the dispersion relations with one subtraction:
\begin{eqnarray}
{\rm Re} f^{(+)}(x)=
\frac{x^2}{\pi}\,v.p.\, \int\limits^{\infty}_{-\infty}dx^\prime 
\frac{{\rm Im} f^{(+)}(x^\prime+i\epsilon^\prime)}{x^{\prime\, 2}(x^\prime - x)}+{\rm Re} f^{(+)}(0),
\nonumber
\end{eqnarray}
where one used $[{\rm Re} f^{(+)}(0)]^\prime x = 0$.

\section{Dispersion relations in skewness-plane \cite{T-An}}

We restrict our study by the case: $\{ s,\, Q^2\} \rightarrow \infty, t \ll \{s,\, Q^2\}$, 
where QCD factorization is applicable.  
At the leading order, DVCS and VMP amplitudes can be expressed via 
\begin{eqnarray}
\label{ampLO}
{\cal A}_{f}(\xi,\,t)=\lim_{\epsilon\to 0} {\cal A}_{f}(\xi-i\epsilon,\,t)=
\lim_{\epsilon\to 0}\int\limits_{-1}^{1} dx\, 
\frac{H^{(+)}_{f}(x,\,\xi,\,t)}{x-\xi+i\epsilon}.
\end{eqnarray} 
Here, $H^{(+)}(x,\,\xi,\,t)$ denotes the  singlet ($C=+1$) combination of GPDs, summing the contributions of 
quarks and anti-quarks and of $s$- and $u$-channels. 

\paragraph{Analyticity of amplitudes}
To prove the analyticity of the amplitude for $|\xi| > 1$ , one represents the denominator as 
the geometric series \cite{T-An}:
\begin{eqnarray} 
\label{decan}
{\cal A}(\xi)=- \sum_{n=0}^{\infty} \xi^{-n-1}
\int\limits_{-1}^{1} dx\,H^{(+)}(x,\,\xi) x^n.
\end{eqnarray}
This series is convergent thanks to the polynomiality condition which reflects the Lorentz invariance \cite{GPV}.
One can write down the fixed-t dispersion relations for the LO amplitude:
\begin{eqnarray}
\label{dr20}
{\rm Re}\,{\cal A}(\xi)=
\frac{v.p.}{\pi }\int\limits_{-1}^{1} dx\, 
\frac{{\rm Im}{\cal A}(x+i\epsilon)}{x-\xi}+\Delta(\xi),
\end{eqnarray}  
where the limit $\epsilon\to 0$ is implied. 
In other words, we write
\begin{eqnarray}
\label{dr2}
v.p.\int\limits_{-1}^{1} dx\, 
\frac{H^{(+)}(x,\,\xi)}{x-\xi}=
v.p.\int\limits_{-1}^{1} dx\, 
\frac{H^{(+)}(x,\,x)}{x-\xi}+\Delta(\xi).
\end{eqnarray} 
Here, $\Delta(\xi)$ is a possible subtraction.
This expression represents the holographic property of GPD:
the relevant information about hard exclusive amplitudes in the considered leading
approximation is contained in the one-dimensional sections $x= \pm \xi$
of the two dimensional space of $x$ and $\xi$. These holographic as well as tomographic
properties in momentum space 
are complementary to the often discussed  holography and tomography in coordinate space.

Prove now that $\Delta(\xi)$ is finite and  independent on $\xi$. 
To this end, one considers the following representation \cite{T-An}:
\begin{eqnarray} 
\label{diff2}
\Delta(\xi)=v.p.\int\limits_{-1}^{1} dx\, 
\frac{H^{(+)}(x,\,\xi)-H^{(+)}(x,\,x)}{x-\xi}= 
-v.p.\int\limits_{-1}^{1} dx\, \sum_{n=1}^{\infty} 
\frac{1}{n!} \frac{\partial^n}{\partial\xi^n}H^{(+)}(x,\,\xi)\biggr|_{\xi=x}(\xi-x)^{n-1}. \nonumber
\end{eqnarray}
Due to the polynomiality condition the only surviving highest power term  
in this series is equal to a finite subtraction constant:
$\Delta(\xi)\equiv\Delta$. This can also be derived with a use of the
Double Distributions (DDs) formalism \cite{T-An, AT}.
It should be emphasized that both integrals in (\ref{dr2})
are divergent at $\xi=t=0$, and this divergencies do not cancel for $\xi \to 0$.
Thus, $\Delta$ can not be defined for $\xi=t=0$. But,
the point $\xi=0$ (or $Q^2=0$)  cannot be accessed in DVCS and VMP experiments.
Note that for an arbitrarily small $\xi$ the integrals are finite and, therefore,
$\Delta$ is well-defined.

Taking into account the parameterization for $D$-term as \cite{GPV}
\begin{eqnarray}
\label{parD}
D(\beta)=(1-\beta^2) \sum_{n=0}^{\infty} d_n C^{(3/2)}_{2n+1}(\beta),
\end{eqnarray}
and keeping only the lowest term, one gets
$\Delta= - 4 d_0$.
This lowest term $d_0$ was estimated within the framework of different models.
We focus on the results of chiral quark-soliton model \cite{CQM}: 
$d_0^{{\rm CQM}}(N_f)=d_0^{u}=d_0^{d}=-\frac{4.0}{N_f}$,
where $N_f$ is the number of active flavours,
and lattice simulations \cite{Lattice}: 
$d_0^{{\rm latt}}=d_0^{u}\approx d_0^{d} =-0.5$.
The subtraction constant varies as $\Delta^{p}_{{\rm CQM}}(2)=\Delta^{n}_{{\rm CQM}}(2)\approx 4.4, \,
\Delta^{p}_{{\rm latt}}\approx\Delta^{n}_{{\rm latt}}\approx 1.1$
for the DVCS on both the proton and neutron targets.

\section{Dispersion relations in $\nu$-plane \cite{AT}}

Let us now compare the dispersion relation 
with the dispersion relation written in $\nu$-plane  where 
$\nu=(s-u)/4m_N$.
In terms of new variables  $\nu^\prime,\, \nu$ related to $x,\, \xi$ as 
$x^{-1}=4m_N\nu^\prime/Q^2,\,\xi^{-1}=4m_N\nu/Q^2 $. 
The fixed-$t$ dispersion relation becomes the subtracted one:
\begin{eqnarray}
\label{drsub}
{\rm Re}\,{\cal A}(\nu, Q^2)=
\frac{v.p.}{\pi}\int\limits_{Q^2/4m_N}^{\infty} d\nu^{\prime\, 2}
{\rm Im}\,{\cal A}(\nu^{\prime}, Q^2)\biggl[ \frac{1}{\nu^{\prime\, 2}-\nu^2} -  
\frac{1}{\nu^{\prime\, 2}}\biggr]+ \Delta.
\end{eqnarray}
This subtracted (in the symmetric unphysical point $\nu=0$)
dispersion relation is our principal result.
It can be considerably simplified provided ${\rm Im}\,{\cal A}(\nu)$ decreases fast enough 
so that both terms in the squared brackets can be integrated separately:  
\begin{eqnarray}
\label{dr1}
{\rm Re}\,{\cal A}(\nu)=
\frac{v.p.}{\pi}\int\limits_{\nu_0}^{\infty} d\nu^{\prime\, 2}\, 
\frac{{\rm Im}\,{\cal A}(\nu^{\prime})}{\nu^{\prime\, 2}-\nu^2} + {\bf C}_0 \, ,
\end{eqnarray}
where 
\begin{eqnarray}
\label{C0}
{\bf C}_0=\Delta -
\frac{1}{\pi}\int\limits_{\nu_0}^{\infty} d\nu^{\prime\, 2}\,  
\frac{{\rm Im}\,{\cal A}(\nu^{\prime})}{\nu^{\prime\,2}}
 = \Delta + \int\limits_{-1}^{1} dx\, 
\frac{H^{(+)}(x,\,x)}{x}.
\end{eqnarray}
Using (\ref{dr2}) with $\xi=0$, 
\begin{eqnarray}
\label{subtr1}
\Delta (t)&=&2 \int\limits_{-1}^{1} dx\, 
\frac{H(x,\,0,t)-H(x,\,x,t)}{x},
\end{eqnarray}
one can see that the $D$-term is cancelled from the expression for the subtraction constant
\begin{eqnarray}
\label{Const}
{\bf C}_0(t)= 2 \int\limits_{-1}^{1} dx 
\frac{H(x,\,0,\,t)}{x}.
\end{eqnarray}
This constant is similar to the result obtained in the studies of the fixed pole 
contribution to the forward Compton amplitude.
For $\xi,\, t=0$, GPDs are expressed in terms of standard parton distributions  
\begin{eqnarray}
{\bf C}_0(0)=2\int\limits_{0}^{1} dx 
\frac{q(x)+\bar q(x)}{x} 
= 2\int\limits_{0}^{1} dx 
\frac{q_v(x)+ 2 \bar q(x)}{x}.
\end{eqnarray}

However, the integrals in (\ref{subtr1}) and (\ref{Const}) diverge at low $x$ in both the valence and sea quark contributions \cite{AT,PolT}.
Therefore, for $t=0$ we should consider (\ref{drsub}) as a correct general form of the dispersion relation which includes
the infinite subtraction at the point $\nu=0$ and the subtraction constant associated with the $D$-term \cite{AT}.  

For $t\neq 0$, the integrals in (\ref{subtr1}) and (\ref{Const}) converge for sufficiently large $t$ \cite{AT}. 
In the case of Regge inspired parameterization \cite{GPV} $H(x,\,0,\,-t)\sim x^{-\alpha(0)+\alpha^\prime t}$,  
this condition reads as $t>\alpha(0)/\alpha^\prime$,
resulting in $t \gtrsim 1 (10){\rm GeV}^{2}$ for the valence (sea)
quark distributions. 

The subtracted dispersion relation for the forward Compton scattering amplitude is
\begin{eqnarray}
\label{drsubF}
&&{\rm Re}\,{\cal A}(\nu)=
\frac{\nu^2}{\pi}v.p.\int\limits_{0}^{\infty} \frac{d\nu^{\prime\, 2}}{\nu^{\prime\, 2}}\,  
\frac{{\rm Im}\,{\cal A}(\nu^{\prime})}{(\nu^{\prime\,2}-\nu^2)} + \Delta.
\end{eqnarray}
One can see that, for the proton target, our subtraction combined with the 
lattice simulations
is rather close to the low energy Thomson term (note that $\Delta_{{\rm Thomson}}=1$ for our normalization of the 
Compton amplitude).

Note that approaching the value $Q^2 =0$ for Compton amplitude involves the infinite tower of higher twists.
The observed numerical coincidence may signal about the sort of duality similar to Bloom-Gilman type duality between 
leading twist contribution and full result \cite{dual}.

\section{Conclusions} 

We can conclude that the fixed-$t$ DR for the DVCS and VMP amplitudes  
require the infinite subtractions at $\nu=0$ with the subtraction 
constants associated with the $D$-terms. 
For the productions of the mesons defined by valence ($C=-1$) GPDs
the finite subtraction is absent.  
The appearance of the subtraction expressed in terms of (forward) parton distributions  
may be investigated in the framework of the leading order QCD factorization. 
The possibility of continuation of our results to the real photons limit has been considered.  
 
{\it Acknowledgments.} We would like to thank A.P.~Bakulev, M.~Diehl, 
 A.V.~Efremov, S.B.~Gerasimov, D.Yu.~Ivanov, S.V.~Mikhailov, D.~ M\"uller, B.~Pire, M.V.~Polyakov, A.V.~Radyushkin, L.~Szymanowski, 
S.~Wallon for useful discussions and correspondence.  
This work was supported in part by Deutsche Forschungsgemeinschaft
(Grant 436 RUS 113/881/0), RFBR (Grants 06-02-16215 and 07-02-91557) and 
RF MSE RNP (Grant 2.2.2.2.6546).

\end{document}